\documentclass[11pt]{article}
\usepackage{amsfonts}

\newcommand{\beq}{\begin{equation}}
\newcommand{\eeq}{\end{equation}}
\newcommand{\ben}{\begin{eqnarray}}
\newcommand{\een}{\end{eqnarray}}
\newcommand{\f}{\frac}

\newcommand{\ds}{\displaystyle}

\def\1{\'{\i}}                           
 
\def\>#1{{\bf #1}}

\def\adl{L}
\def\tt{T}

\def\gadl{\hat l}
\def\gtt{\hat t}
\def\radio{R}

\begin{document}
 
\baselineskip=18pt

\thispagestyle{empty}

\ 
\hfill\

\begin{center}

{\Large{\bf{A  non-commutative Minkowskian
spacetime from a quantum AdS algebra}}}

\end{center}

\bigskip 
\medskip

\begin{center}   
Angel~BALLESTEROS$^a$, N.~Rossano~BRUNO$^{a,b}$, Francisco~J.~HERRANZ$^a$
\end{center}

\begin{center} {\it { 
${}^a$Departamento de F\1sica,
Universidad de Burgos, Pza.\ Misael Ba\~nuelos s.n., \\
09001 Burgos, Spain }}\\
e-mail: angelb@ubu.es, fjherranz@ubu.es
\end{center}

\begin{center} {\it { 
${}^b$Dipartimento di Fisica, Universit\`a di Roma ROMA TRE, and INFN Sez.\ Roma Tre, Via Vasca Navale 84,
00146 Roma, Italy}}\\
e-mail: rossano@fis.uniroma3.it
\end{center}

\bigskip 
\medskip

\begin{abstract}
\noindent
A   quantum deformation of the conformal algebra of the  Minkowskian
spacetime in $(3+1)$ dimensions is identified with a deformation of  the
$(4+1)$-dimensional AdS algebra.   Both  Minkowskian
and AdS first-order non-commutative spaces are explicitly obtained, and the former coincides with
the well known $\kappa$-Minkowski space. Next, by working in the conformal basis, a   new
non-commutative Minkowskian spacetime is constructed through the full (all orders) dual quantum
group spanned by deformed Poincar\'e and dilation symmetries. Although Lorentz invariance is
lost,  the resulting   non-commutative spacetime  is quantum group covariant, preserves 
  space isotropy and, furthermore, can be interpreted as a  generalization of      the  
$\kappa$-Minkowski space in which a   variable fundamental scale (Planck length)  
appears.
\end{abstract}

\bigskip\medskip 

\noindent
KEYWORDS: quantum algebras,  
deformation, Minkowski, Anti-deSitter, Poincar\'e, non-commutative spacetime

\noindent
PACS:  02.20.Uw\quad 11.30.-j\quad 04.60.-m

\bigskip\medskip 



\newpage

\section{Introduction}

One of the most relevant applications of quantum groups in physics is  
the construction of deformed spacetime symmetries  that generalize    classical
Poincar\'e kinematics beyond   Lie algebras,
such as   the well known $\kappa$-Poincar\'e
\cite{LukierskiRuegg1992,Giller,Lukierskib} and the quantum null-plane (or light-cone)
Poincar\'e  
\cite{nulla,nullb} algebras. For  all these cases, the deformation parameter has been interpreted
as a fundamental scale which may be related with the Planck length. In fact, these results 
can be seen as   different attempts to  develop new approaches to physics at the Planck
scale, an idea that was early presented in \cite{Majida}. A further physical development of  the
$\kappa$-Poincar\'e   algebra   has
  led to the so called  doubly special relativity (DSR) theories
\cite{Amelino-Camelia:2001xx,Amelino-Camelia:2000mn,
MagueijoSmolin,Amelino-Camelia:2002vy,Kowalski-Glikman:2002we}
that analyse the   fundamental role assigned to the deformation parameter/Planck length as an
observer-independent length scale to be considered together with the usual observer-independent
velocity scale $c$, in such a manner that  Lorentz invariance is preserved
\cite{Bruno:primo,Bruno:2002wc,Lukierski:2002df}.

From a dual quantum group  perspective, when the quantum
spacetime coordinates $\hat x^\mu$ conjugated to the  $\kappa$-Poincar\'e momentum-space
$P_\mu$ (translations)
 are considered, the        non-commutative $\kappa$-Minkowski spacetime 
arises \cite{Maslanka,Majid:1994cy,Zak,LukR}. Some    field theories on such a space have been
proposed (see \cite{Amelino-Camelia:2002mu} and
 references therein) and its role in  DSR theories has been analysed
\cite{Kowalski-Glikman:2002jr}. More general   non-commutative Minkowskian
spacetimes can be expressed  by means of the following Lie algebra commutation
rules~\cite{Lukierskid}: 
\beq
[\hat x^\mu, \hat x^\nu]=\frac 1{\kappa} (a^\mu\hat x^\nu-a^\nu\hat x^\mu),
\label{aa}
\eeq
where $a^\mu$ is a {\em constant} four-vector in the  Minkowskian space.

In the context of $\kappa$-deformations, which are understood as quantum algebras with a
dimensionful deformation parameter related with the Planck length,  Poincar\'e  symmetry should be
taken only as a first stage that should be  embedded in some way within    more general structures
such as deformed conformal or AdS/dS symmetries, that is, quantum $so(4,2)$/$so(5,1)$ algebras.
Thus, it is natural to think that  a non-commutative  Minkowskian space of the form (\ref{aa})
could also be either embedded or generalized. In this respect, by considering the simplest quantum
deformation of the Weyl--Poincar\'e algebra (isometries plus dilations), $U_\tau({\cal WP})$
\cite{Herranz:2002fe}, a  new DSR proposal  has been presented in  \cite{Ball}. Such a quantum
algebra   arises as a Hopf subalgebra of  a `mass-like' quantum deformation $U_\tau(so(4,2))$ of
the   conformal algebra of the $(3+1)$D Minkowskian space.

The aim of   this letter is  to analyse the   first-order (in both the deformation parameter
and non-commutative coordinates) quantum group dual to 
$U_\tau(so(4,2))$, to
construct   the  complete (all orders) Hopf algebra dual to
$U_\tau({\cal WP})$  and, afterwards,    to extract some physical implications
of the associated non-commutative Minkows\-kian spacetime.

In the next section we  identify the
 deformation $U_\tau(so(4,2))$, formerly obtained  in a conformal basis, with a   $(4+1)$D quantum
AdS algebra at both algebra and dual   group levels. By using the Hopf subalgebra
spanned by the deformed
Poincar\'e and dilation generators   we compute in
section 3  the associated dual quantum group by making use of the quantum $\cal R$-matrix. The
last  section is devoted to derive the physical consequences conveyed by the resulting
non-commutative Minlowskian spacetime which  is covariant under quantum group transformations and
does preserve space isotropy, although its Lorentz invariance is lost. Moreover,  this  new
non-commutative spacetime generalizes $\kappa$-Minkowski space since it is defined through Lie
algebra commutation rules whose structure constants are  the quantum group entries associated to
the Lorentz sector.

\section{Quantum  AdS algebra}

Let us consider the   quantum deformation \cite{Herranz:2002fe}  of the conformal 
algebra of the $(3+1)$D Minkowskian  spacetime, $U_\tau(so(4,2))\equiv
U_\tau({\cal CM}^{3+1})$,  
  which is spanned by
the  generators of rotations $J_i$, time and space
translations   $P_\mu$, boosts $K_i$,   conformal transformations $C_\mu$ and
dilations $D$. Hereafter we   assume    $c= \hbar =1$ ($\omega=1$ in  \cite{Herranz:2002fe}), 
 sum over repeated indices,  Latin indices $i,j,k=1,2,3$, and Greek
indices $\mu,\nu=0,1,2,3$. The  non-vanishing deformed commutation rules and coproduct 
for $U_\tau({\cal CM}^{3+1})$ are given by
 \beq
\begin{array}{lll}
[J_i,J_j]=   \varepsilon_{ijk}J_k ,&\qquad
[J_i,K_j]=   \varepsilon_{ijk}K_k ,&\qquad
[J_i,P_j]=  \varepsilon_{ijk}P_k , \\[2pt]
[K_i,K_j]=-  \varepsilon_{ijk}J_k ,
 &\qquad   [K_i,P_0]= {\rm e}^{-\tau P_0} P_i ,&\qquad
 [D,P_i]= P_i ,\\[2pt]
   \displaystyle{ [K_i,P_i]=   
   \frac{{\rm e}^{\tau P_0}-1}{\tau} } ,&\qquad
\displaystyle{[D,P_0]=  \frac{1-{\rm e}^{-\tau P_0}}{\tau}} ,&
\end{array}
\label{ba}
\eeq  
\beq
\begin{array}{lll}
[J_i,C_j]= \varepsilon_{ijk}C_k  ,&\qquad 
 [D,C_i]=-  C_i ,&\qquad   [D,C_0]=-   C_0+\tau D^2   , \\[2pt]
[P_0,C_0]=-2    D , &\qquad 
[K_i,C_0]=  C_i  ,&\qquad  [K_i,C_i]=   C_0 -\tau D^2  , \\[2pt]
\multicolumn3{l}{ [C_0,C_i]=-  \tau(D C_i + C_i D) ,
 \qquad\quad [P_0,C_i]=    {\rm e}^{-\tau P_0}
K_i+K_i\,{\rm e}^{-\tau P_0}   ,}\\[2pt] 
\multicolumn3{l}{  [P_i,C_j]=2  (\delta_{ij}D    -    \varepsilon_{ijk} J_k) ,
\qquad\quad
 [C_0,P_i]=2  K_i+ \tau (D P_i + P_i D)  , } 
  \end{array}
\label{bb}
\eeq

\beq
\begin{array}{l}
\Delta(P_0)=1\otimes P_0 + P_0\otimes 1  ,
\qquad \Delta(P_i)=1\otimes P_i + P_i\otimes {\rm e}^{\tau P_0}  ,\\[4pt]
\Delta(J_i)= 1\otimes J_i + J_i\otimes 1 ,\qquad\ \ \Delta(D)=1\otimes D +
D\otimes {\rm e}^{-\tau P_0} ,\\[4pt]
\Delta(K_i)=1\otimes K_i + K_i\otimes 1-\tau D\otimes  {\rm
 e}^{-\tau P_0}P_i  ,\qquad \Delta(C_0)=1\otimes C_0 + C_0\otimes {\rm
e}^{-\tau P_0}  ,\nonumber\\[4pt] 
\Delta(C_i)=1\otimes C_i + C_i\otimes {\rm e}^{-\tau
P_0}+2\tau D\otimes  {\rm e}^{-\tau P_0}K_i-\tau^2 (D^2+ D) \otimes  {\rm
e}^{-2\tau P_0} P_i .
 \end{array} 
 \label{bc}
\eeq
The  
deformation parameter  $\tau$   will be   identified with the   Planck
length:  $\tau\sim  L_p$ \cite{Ball}.

Some properties of    $U_\tau(so(4,2))$ can be unveiled by studying its Lie bialgebra  structure,
that is, the first-order relations in the
 deformation parameter,  generators and dual coordinates. If we write the deformed coproduct
$\Delta$  as   formal power series in  $\tau$,   
\beq
\Delta=\sum_{k=0}^{\infty} \Delta_{(k)}=\sum_{k=0}^{\infty}
\tau^k \delta_{(k)}, 
\eeq
the cocommutator $\delta$ is given by the skewsymmetric part of the first-order deformation,
\beq
  \delta =\delta_{(1)} - \sigma\circ
\delta_{(1)},
\label{coco}
\eeq
where  $\sigma(a\otimes b)=b\otimes a$. In our case, from (\ref{bc}) we find that
\beq
\begin{array}{l}
\delta(P_0)=0,\qquad \delta(P_i)=\tau P_i\wedge P_0,\\
\delta(J_i)=0,\qquad \delta(D)=-\tau D\wedge P_0,\\
\delta(K_i)=-\tau D\wedge P_i,\qquad \delta(C_0)=-\tau C_0\wedge P_0,\\
\delta(C_i)=-\tau C_i\wedge P_0+2\tau D\wedge K_i,
 \end{array} 
 \label{cocob}
\eeq
where $\wedge$ denotes the skewsymmetric tensor product. Next, 
  Lie bialgebra     duality \cite{Majid,CP}  leads to the dual commutation rules in the form
\beq
\delta(Y_i)=f_i^{jk}Y_j\wedge Y_k \quad \Rightarrow\quad 
[\hat y^j,\hat y^k]=f_i^{jk}\hat y^i ,
\label{dual}
\eeq
where $Y_i$ is a generic generator and $\hat y^i$ its associated dual quantum group coordinate
fulfilling
$\langle \hat y^i| Y_j\rangle=    \delta_j^i$. 
Therefore if we denote by  $\{ \hat
x^\mu, \hat\theta^i,\hat\xi^i,\hat d,\hat c^\mu\} $ the dual non-commutative coordinates of the 
generators $\{P_\mu,J_i,K_i,D,C_\mu\}$, respectively, we obtain  from (\ref{cocob}) the
following  non-vanishing first-order quantum group commutation rules:
\beq
\begin{array}{lll}
[\hat x^0,\hat x^i]=-\tau \hat x^i, &\qquad
[\hat x^0,\hat d]=\tau \hat d, &\qquad
[\hat x^0,\hat c^\mu]=\tau \hat c^\mu, \\[4pt]
[\hat d,\hat x^i]=-\tau \hat \xi^i, &\qquad
[\hat d,\hat \xi^i]=2 \tau \hat c^i . & 
 \end{array} 
 \label{bd}
\eeq
The   non-commutative
  Minkowskian spacetime is then characterized by 
\beq
[\hat x^0,\hat x^i]=-\tau
\hat x^i,\qquad [\hat x^i,\hat x^j]=0,
\label{bbdd}
\eeq
 which coincides with the  usual $\kappa$-Minkowski space (\ref{aa}) with   $a^\mu=(1,0,0,0)$, 
provided that $\tau=-1/\kappa$.  Nevertheless,  in the
next  section we shall compute the  full (all orders) dual quantum  group to $U_\tau({\cal WP})$,
and the associated
  non-commutative spacetime will generalize  the first-order relations (\ref{bbdd}).

We stress that all the above expressions can be rewritten in terms of deformed symmetries
of the $(4+1)$D AdS spacetime, $AdS^{4+1}$.   Let $\adl_{AB}$ $(A<B)$ and $\tt_{A}$
($A,B=0,1\dots,4$) be the Lorentz  and   translations undeformed generators  obeying
\beq
\begin{array}{l}
[\adl_{AB},\adl_{CD}]=\eta_{AC}\adl_{BD}-\eta_{AD}\adl_{BC}-\eta_{BC}\adl_{AD}
+\eta_{BD}\adl_{AC},\\[4pt]
[\adl_{AB},\tt_{C}]=\eta_{AC}\tt_{B}-\eta_{BC}\tt_{A},\qquad [\tt_A,\tt_B]=-\frac
1{\radio^2} \adl_{AB},
 \end{array} 
 \label{be}
\eeq
such that $\eta=(\eta_{AB})={\rm diag}\,(-1,1,1,1,1)$ is the Lorentz metric
associated to $so(4,1)$, $\adl_{0B}$ are the four  boosts in $AdS^{4+1}$ and $R$ is the AdS
radius related with the cosmological constant by
$\Lambda=6/R^2$.
Then  the following change of basis ($i=1,2,3$):
\beq
\begin{array}{lll}
\tt_0=-\frac 1{2\radio} (C_0+P_0),&\quad \tt_1=\frac 1{ \radio}\, D,&\quad \tt_{i+1}=\frac
1{2\radio}(C_i+P_i),\\[4pt]
\adl_{01}=\frac 12 (C_0-P_0),&\quad \adl_{0,i+1}=K_{i},&\quad \adl_{1,i+1}=\frac
12(C_{i}-P_i),\\[4pt]
\adl_{23}=J_3,&\quad \adl_{24}=-J_2,&\quad \adl_{34}=J_1, 
 \end{array} 
 \label{bf}
\eeq
connects ${\cal CM}^{3+1}$ with $AdS^{4+1}$ and can be taken in the deformed case as a way to
identify  (\ref{ba})--(\ref{bc}) as the quantum deformation  $U_\tau(so(4,2))\equiv
U_\tau( AdS^{4+1})$.
The   dual Lorentz ${\gadl}^{AB}$ and  
spacetime ${\gtt}^A$ quantum AdS-coordinates  can also be written in terms of
the conformal ones as 
\beq
\begin{array}{lll}
\gtt^0=- R (\hat c^0+\hat x^0),&\quad \gtt^1=R\, \hat d,&\quad \gtt^{i+1}= R (\hat
c^{i}+\hat x^{i} ),\\[4pt]
\gadl^{01}=\hat c^0-\hat x^0 ,&\quad \gadl^{0,i+1}=\hat \xi^{i},&\quad \gadl^{1,i+1}=
\hat c^{i}-\hat x^{i} ,\\[4pt]
\gadl^{23}=\hat \theta^3,&\quad \gadl^{24}=-\hat \theta^2,&\quad \gadl^{34}=\hat
\theta^1 . 
 \end{array} 
 \label{bg}
\eeq
Hence  the  (first-order) non-vanishing  commutation rules    for the non-commutative $AdS^{4+1}$
spacetime   turn  out to be
\beq
[\gtt^0,\gtt^1]=-   \tau  R\,\gtt^1,\qquad
[\gtt^0,\gtt^{i+1}]=-\tau R^2  \gadl^{1,i+1},\qquad
[\gtt^1,\gtt^{i+1}]=-\tau R^2  \gadl^{0,i+1},
 \label{bbgg}
\eeq
which involve both  the boost $\gadl^{0,i+1}$ and the rotation $\gadl^{1,i+1}$ quantum
coordinates. A   Lie bialgebra contraction analysis \cite{LBC} shows that the  
contraction $R\to \infty$  from
$U_{\tau}( AdS^{4+1})$  and its dual   to a
$(4+1)$D quantum Poincar\'e algebra/group  is well defined whenever   the deformation
parameter is transformed as $\tau\to  \tau R^2$.

Therefore, the maps (\ref{bf}) and (\ref{bg})  can be used   to express 
 the same quantum deformation of
$so(4,2)$  within two physically different
frameworks: $U_\tau( {\cal CM}^{3+1})\leftrightarrow  U_{\tau}( AdS^{4+1})$. In fact, such
a quantum group relationship might  further be applied in order to analyse the role that quantum
deformations of $so(4,2)$ could play  in relation  with   the   ``AdS-CFT  correspondence" that
relates local QFT on
$AdS^{(d-1)+1}$ with a conformal QFT on the  (compactified) Minkowskian
spacetime 
${\cal CM}^{(d-2)+1}$  \cite{Maldacena, Witten,Reh} (in our case up to $d=5$).
We also remark that  a more general (three-parameter) quantum deformation 
of $o(3,2)$ can be found in \cite{Moz}, where the  connection between the corresponding
quantum  ${\cal CM}^{2+1}$   and   $AdS^{3+1}$ algebras is explicitly described.

\section{Quantum  Weyl--Poincar\'e group}

The classical $r$-matrix associated to    $U_\tau(so(4,2))$  reads
\beq
\begin{array}{l}
 r=-\tau
D\wedge P_0\equiv \tau R^2\tt_1\wedge \tt_0+\tau R\, \tt_1\wedge
\adl_{01},
\end{array} 
\label{bh}
\eeq
which  satisfies the classical Yang-Baxter
equation  \cite{Dr}. Therefore, $U_\tau(so(4,2))$ is   a triangular (or
twisting) quantum deformation, different to the Drinfeld--Jimbo type, which 
is supported by the Hopf subalgebra spanned by $\{D,P_0\}$. The 
universal $\cal T$-matrix for the latter can be written as~\cite{Tmatrix}
\beq
{\cal T}={\rm e}^{\hat d D}
\,{\rm e}^{\hat x^0 P_0},
\label{t2}
\eeq 
 while     the  ${\cal R}$-matrix reads
\beq
\mathcal{R}=\exp(\tau P_0\otimes D) \exp (-\tau D\otimes P_0) .
\label{da}
\eeq
Since this element is also a universal ${\cal R}$-matrix for both $U_\tau({\cal
WP})\subset  U_\tau(so(4,2))$ \cite{Herranz:2002fe},   the
corresponding  dual quantum groups can be deduced explicitly by applying
the FRT procedure~\cite{Faddeev:ih}. This requires a   matrix 
representation $R$ for
(\ref{da}) as well as   to choose a matrix element  $T$ of the quantum group 
with non-commutative entries. However,  the   consideration    of the
complete 
$U_\tau(so(4,2))$ structure (in both   conformal and AdS bases) precludes a clear
identification of the non-commutative spacetime  coordinates as these would appear
as arguments of functions that also depend on many other    coordinates. In this
respect  see, for instance, \cite{Chang} for the construction of a quantum AdS space
from a $q$-$SO(3,2)$ of Drinfeld--Jimbo type.

Since  we are mainly interested in the structure and physical consequences of the associated
non-commutative spacetime coordinates (dual to the translation generators),  we    shall
consider here the FRT construction  for the Weyl--Poincar\'e  Hopf subalgebra in the
conformal basis.

A deformed matrix representation for the  defining relations of $U_\tau({\cal WP})$  
(\ref{ba})  can be obtained from the $6\times 6$   matrix representation of $AdS^{4+1}$; namely,
\beq
\begin{array}{l}
P_0=\f{\tau}{2} (e_{00}- e_{01}+ e_{10}-
 e_{11})-e_{02}-e_{12}+e_{20}-e_{21},\\[4pt]
 P_i=e_{0,i+2}+e_{1,i+2}+e_{i+2,0}-e_{i+2,1},\qquad    D= e_{01}+e_{10},\\[4pt]
 J_i=- \varepsilon_{ijk}\, e_{j+2,k+2} ,\qquad
 K_i=e_{2,i+2}+e_{i+2,2}, 
\end{array} 
\label{db}
\eeq
where $e_{ab}$ ($a,b=0,\dots,5$) is the matrix with entries $\delta_{ab}$. 
We construct the quantum group element $T$ in such a representation by considering
 the following    matrix product, which is consistent with the exponential form of the
universal $\cal T$-matrix (\ref{t2}) for the carrier subalgebra $\{D,P_0\}$:  
\beq
\begin{array}{ll}
{T}\!\!\! &={\rm e}^{\hat d D}
\,{\rm e}^{\hat x^0 P_0} 
\,{\rm e}^{\hat x^1 P_1}  
\,{\rm e}^{\hat x^2 P_2}  
\,{\rm e}^{\hat x^3 P_3} 
\,{\rm e}^{\hat \theta^1 J_1}  
\,{\rm e}^{\hat \theta^2 J_2}  
\,{\rm e}^{\hat \theta^3 J_3} 
\,{\rm e}^{\hat \xi^1 K_1}  
\,{\rm e}^{\hat \xi^2 K_2}  
\,{\rm e}^{\hat \xi^3 K_3}  \\[4pt]
& = \left(\begin{array}{cccccc}
\hat\alpha_+&\hat\beta_-&\hat\gamma_0&
\hat\gamma_1&\hat\gamma_2&\hat\gamma_3\\
\hat\beta_+&\hat\alpha_-&\hat\gamma_0&
\hat\gamma_1&\hat\gamma_2&\hat\gamma_3\\
\hat x^0&- \hat x^0&\hat\Lambda^0_0&
\hat\Lambda^0_1&\hat\Lambda^0_2&\hat\Lambda^0_3\\
\hat x^1&- \hat x^1&\hat\Lambda^1_0&
\hat\Lambda^1_1&\hat\Lambda^1_2&\hat\Lambda^1_3\\
\hat x^2&- \hat x^2&\hat\Lambda^2_0&
\hat\Lambda^2_1&\hat\Lambda^2_2&\hat\Lambda^2_3\\
\hat x^3&- \hat x^3&\hat\Lambda^3_0&
\hat\Lambda^3_1&\hat\Lambda^3_2&\hat\Lambda^3_3
\end{array}\right) ,
\end{array}
\label{dd}
\eeq
where the non-commutative entries are the quantum Minkowskian coordinates $\hat x^\mu$ and
\beq
\begin{array}{l}
\hat\alpha_\pm=\cosh\hat d\pm \frac 12{\rm e}^{\hat d}(\hat x_\mu\hat x^\mu
+\tau\hat x^0) ,\qquad \hat\gamma_\nu={\rm e}^{\hat d} \hat x_\mu
\hat \Lambda^\mu_\nu,\\[4pt]
\hat\beta_\pm=\sinh\hat d\pm \frac 12{\rm e}^{\hat d}(\hat x_\mu\hat x^\mu
+\tau\hat x^0) ,\qquad \hat\Lambda^\mu_{\nu}
=\hat\Lambda^\mu_{\nu}(\theta^i,\xi^i),\\[4pt]
\hat\Lambda^\mu_{\nu} \hat\Lambda^\rho_{\sigma} g^{\nu\sigma}
=g^{\mu\rho},\qquad  \hat x_\mu=g_{\mu\nu} \hat x^\nu,\qquad
(g^{\mu\rho})={\rm diag}\,(-1,1,1,1) .
\label{de}
\end{array}
\eeq
Note that  quantum rotation  and boost coordinates are jointly expressed through 
the formal   Lorentz entries $\hat\Lambda^\mu_{\nu}$.

   The representation
(\ref{db}) gives rise to  a quantum ${R}$-matrix
(\ref{da}) with dimension  $36\times
36$. Since $P_0^3$ vanishes,  ${R}$ reads
\beq
\begin{array}{l}
{R}=( \mbox{\boldmath $1$}\otimes \mbox{\boldmath $1$}+\tau P_0\otimes D
+\frac 12\tau^2 P_0^2\otimes D^2)
( \mbox{\boldmath $1$}\otimes \mbox{\boldmath $1$}-\tau D\otimes P_0
+\frac 12\tau^2 D^2\otimes P_0^2)  ,
\end{array}
\label{df}
\eeq
where $\mbox{\boldmath $1$}$ is the $6\times 6$ unit  
matrix. Next in pursuing the FRT program we impose that
\beq
{   R}{  T}_1{  T}_2={  T}_2{  T}_1{  R} ,
\label{dg}
\eeq
where ${  T}_1={  T}\otimes \mbox{\boldmath $1$}$ and ${ 
T}_2=\mbox{\boldmath
$1$}\otimes {  T}$. This matrix equation provides the commutation rules among all the entries
in (\ref{dd}), which by taking into account (\ref{de}) can then be  reduced to 
\beq
\begin{array}{l}
 [\hat d,\hat\Lambda^\mu_{\nu} ]=0,\qquad
 [\hat x^\alpha,\hat\Lambda^\mu_{\nu} ]=0,\qquad
[\hat \Lambda^\alpha_{\beta},\hat\Lambda^\mu_{\nu} ]=0, \\[4pt] 
 [\hat d,\hat x^\mu]=\tau\left(
\delta^\mu_{0}{\rm e}^{-\hat d}-\hat \Lambda^\mu_{0} \right),\qquad 
[\hat x^\mu , \hat x^\nu ] =\tau
  \left(\hat \Lambda^\nu_{0}\hat x^\mu -
\hat \Lambda^\mu_{0} \hat x^\nu   \right) ,
\end{array}
\label{dh}
\eeq
where the quantum   Lorentz entries  $\hat\Lambda^\mu_{0}$  are given by
\beq
\begin{array}{l}
\hat \Lambda^0_0=\cosh\hat\xi^1\cosh\hat\xi^2\cosh\hat\xi^3 ,
\qquad  \hat\Lambda^2_0= \sinh\hat\xi^2\cosh\hat\xi^3  ,\\[4pt]
 \hat\Lambda^1_0=\sinh\hat\xi^1\cosh\hat\xi^2\cosh\hat\xi^3   ,
\qquad  \hat\Lambda^3_0= \sinh\hat\xi^3 .
\end{array} 
\label{ffa}
\eeq  
The   coproduct for all the entries in ${  T}$ is just $\ds \Delta({ 
T})={  T}\dot\otimes {  T}$.  By using again  (\ref{de}) the coproduct for $\{\hat d,\hat
x^\mu, \hat\Lambda^\mu_{\nu} \}$ can consistently be found:
\beq
\Delta(\hat d)=\hat d\otimes 1+ 1\otimes \hat d,
\quad\ \Delta (\hat x^\mu)=\hat x^\mu \otimes {\rm e}^{-\hat d} +  
\hat \Lambda^\mu_{\eta}\otimes \hat x^\eta
,\quad\ \Delta({\hat \Lambda^\mu_{\nu}})= 
\hat \Lambda^\mu_{\eta} \otimes \hat \Lambda^\eta_{\nu}, 
\label{di}
\eeq
which is a homomorphism of (\ref{dh}).
Thus the expressions (\ref{dh})--(\ref{di})
together with the counit $\epsilon({  T})=
 \mbox{\boldmath $1$}$   and antipode    $S({  T})={ 
T}^{-1}$ determine
 the Hopf algebra structure of the quantum Weyl--Poincar\'e group dual to $U_\tau({\cal
WP})$, which is  the restriction $\hat c^\mu\equiv 0$ of that    dual to $U_\tau( {\cal
CM}^{3+1})$.

We remark that the commutation relations   (\ref{bd}) (with $\hat c^\mu\equiv 0$) can
  be recovered from (\ref{dh}) by only taking the first-order in all the quantum
coordinates (notice that in this case, 
$\hat \Lambda^0_0\to 1$ and $\hat\Lambda^i_0\to \hat\xi^i$).

Recall that the FRT approach was also used in the construction of  the null-plane quantum
Poincar\'e group \cite{nullb}. However,   as  $\kappa$-Poincar\'e has no   universal $\cal
R$-matrix, the  associated quantum group  was obtained   \cite{Maslanka,Majid:1994cy,Zak,LukR} 
through a direct quantization of the semiclassical  Poisson--Lie algebra coming from the
$\kappa$-Poincar\'e  classical $r$-matrix.

\section{A new non-commutative Minkowskian spacetime}

Now we  focus our attention on some structural physical consequences of   the new non-commutative
spacetime that comes out from  (\ref{dh}) 
\beq
[\hat x^\mu , \hat x^\nu ] =\tau
  \left(\hat \Lambda^\nu_{0}(\hat\xi) \hat x^\mu -
\hat \Lambda^\mu_{0}(\hat\xi) \hat x^\nu   \right) ,
\label{fa}
\eeq  
which  can be seen as a   generalization of (\ref{aa}) through   $a^\mu\to \hat
\Lambda^\mu_{0}(\hat\xi)$.

Firstly, we stress that since $\hat \Lambda^\mu_{0}$ (\ref{ffa})  only depend on the quantum
boost parameters and  the
quantum rotation coordiantes $\hat\theta^i$ do not play any role in the spacetime
non-commutativity, the isotropy of the space is thus preserved.
Furthermore, by taking into account the commutation rules   (\ref{dh}), 
  $\hat \Lambda^\mu_{0}$   can be
considered to play the role of the structure
constants within the quantum space (\ref{fa}). In fact, the quantum boost coordinates $\hat\xi^i$  can
be regarded as scalars (usual commutative parameters)  within the quantum
Weyl--Poincar\'e group.  From this viewpoint, relations (\ref{fa}) would define  a Lie algebraic
 non-commutative spacetime of the type (\ref{aa}).  However,   this situation changes   in the
full quantum conformal  group since,  as shown in (\ref{bd}), 
$\hat\xi^i$ and $\hat \Lambda^\mu_{0}$   no  longer commute with the dilation parameter $\hat
d$.

Secondly, relations (\ref{fa})    show that  different observers in
relative motion with respect to quantum group transformations have a different perception of the
spacetime non-commutativity, {\em i.e.},   Lorentz invariance is lost.  Nevertheless, we
remark that, in this context, covariance under quantum group transformations is ensured by
construction. Explicitly, in the
commutative case  the $  T$-matrix (\ref{dd}) is just a matrix representation of the
transformation group of the spacetime, and the   coproduct (\ref{di}) represents the
multiplication of two different elements of the group. A similar  interpretation holds
in the quantum group case, for which (\ref{di}) provides the transformation
law for the non-commutative spacetime coordinates that can be rewritten as
\beq
\hat x''^\mu=\hat x^\mu \, {\rm e}^{-\hat d'} +   \hat\Lambda^\mu_{\,\eta} \,
\hat x'^\eta ,
\label{fb}
\eeq 
where the tensor product notation has been replaced by two different copies of the non-commutative
coordinates  ($\hat x\otimes 1\equiv \hat x$, $1\otimes \hat
x\equiv\hat x'$). The fact that the spacetime (\ref{fa}) is quantum group covariant
is a direct consequence of the Hopf algebra structure which implies that
\beq
[\hat x''^\mu ,\hat x''^\nu ] =\tau \left ( \hat\Lambda''^\nu_{\ 0}(\hat\xi'') \hat
x''^\mu - \hat\Lambda''^\mu_{\ 0}(\hat\xi'') \hat x''^\mu  \right) ,
\label{fc}
\eeq
where the new   Lorentz entries are also given by (\ref{di}):
\beq
  \hat\Lambda''^\mu_{\ \nu}(\hat\xi'')=  \hat\Lambda^\mu_{\, \eta}(\hat\xi) \, \hat
\Lambda'^\eta_\nu (\hat\xi').
\label{fd}
\eeq
Therefore, if we assume that two ``observers" are actually related through a quantum group
transformation (\ref{fb}), they  will be ``affected" by different structure constants for the
spacetime commutation rule (\ref{fa}), yet the latter is manifestly quantum group covariant.

In order to illustrate these results, 
let us consider the  $(1+1)$D case where the   Lorentz sector
is expressed in terms of a single   quantum boost parameter $\hat\xi\equiv \hat\xi^1$
as
\beq
(\hat\Lambda^\mu_{\nu})= \left (
\begin{array}{cc}
\cosh\hat\xi & \sinh\hat\xi\\
\sinh\hat\xi & \cosh\hat\xi
\end{array}
\right ) .
\label{fe}
\eeq
The transformation (\ref{fd})  shows directly the additivity of the quantum boost parameter
(along the same direction):
\beq
\hat\Lambda^0_0(\hat \xi'')=\cosh\hat\xi''=\Delta(\hat\Lambda^0_0)
=\cosh\hat\xi\otimes \cosh\hat\xi+\sinh\hat\xi\otimes \sinh\hat\xi
=\cosh(\hat\xi+\hat\xi'),
\label{ffe}
\eeq
and similarly for the remaining $\hat\Lambda^\mu_{\nu}$. The non-commutative
$(1+1)$D spacetime reads
\beq
[\hat x^1 ,\hat x^0 ] =\tau (\hat x^1   \cosh\hat \xi  -\hat x^0  \sinh\hat
\xi  ) ,
\label{ff} 
\eeq
which implies the following uncertainty relation:
\beq
\delta \hat x^1\, \delta \hat x^0\geq \f{|\tau|}{2}
\left|\langle{\hat x^1}\rangle  \cosh\hat \xi  - \langle{\hat x^0}\rangle 
\sinh\hat\xi  \,\right|,\label{uncert}
\eeq
where $\delta$ denotes  the   root-mean-square deviation and 
$\langle{\hat x^1}\rangle$, $\langle{\hat x^0}\rangle$   are the expectation  values
of the space and time operators.

To end with, we also stress that if      the following new space variables $\hat X^i$    in the
$(3+1)$D spacetime (\ref{fa}) are considered
\beq
\hat x^0 \rightarrow  \hat x^0  ,
\qquad
 \hat x^i \rightarrow   \hat X^i=
\hat x^i\hat \Lambda^0_0-\hat x^0\hat\Lambda^i_0   ,
\eeq
the   transformed commutation rules for the quantum spacetime are given by
\beq
[\hat X^i,\hat x^0]=\tau \hat\Lambda^0_0(\hat\xi)\hat X^i,\qquad [\hat X^i,\hat
X^j]=0,\label{newcoor}
\eeq 
which, in turn,   can be interpreted as a generalization of the $\kappa$-Minkowski space
(\ref{bbdd}) with a ``variable" Planck length $\tau'=\tau \hat\Lambda^0_0(\hat\xi)$ that does
depend on {\em all} the quantum boost parameters (in the $(1+1)$D case, this yields $\tau'=\tau
\cosh\hat\xi$). This result is a direct consequence  of imposing   a larger quantum group
symmetry than Poincar\'e. Moreover, if the quantum conformal transformations and
parameters  are taken into account  and the corresponding quantum
group is constructed, then $\hat \Lambda^0_0$ becomes a non-central operator in such a manner that
(\ref{newcoor}) (and also (\ref{fa})) defines a quadratic non-commutative spacetime. This suggests
that a further study of non-Lie spacetime algebras derived from    conformal or AdS
quantum symmetries could be meaningful.

\section*{Acknowledgements}

This work was partially supported  by MCyT and JCyL,  Spain
(Projects BFM2000-1055 and BU04/03), and by INFN-CICyT
(Italy-Spain).


\end{document}